\documentclass[aps,pra]{revtex4}
\usepackage{graphicx}

\begin{document}

\title{Many-agent controlled teleportation of multi-qubit quantum information
\thanks{Email: zhangzj@wipm.ac.cn}}

\author{Zhan-jun Zhang$^{1,2,*}$ and Zhong-xiao Man$^2$  \\
{\normalsize $^1$ School of Physics \& Material Science, Anhui University, Hefei 230039, China} \\
{\normalsize $^2$ Wuhan Institute of Physics and Mathematics,
Chinese Academy of Sciences, Wuhan 430071, China} \\
{\normalsize *Corresponding author's email address:
zhangzj@wipm.ac.cn}}

\date{\today}
\maketitle

\begin{minipage}{420pt}
We present a general idea to construct methods for multi-qubit
quantum teleportation between two remote parties with control of
many agents in the network. Our methods seem to be much simpler
than the existing method proposed recently (Phys. Rev. A {\bf 70},
022329(2004)). We then demonstrate our idea by using several
different protocols of quantum key distribution, including  Ekert
91 and the deterministic secure communication protocol raised by
Deng and Long.\\

\end{minipage}

No-cloning theorem forbids a perfect copy of an arbitrary unknown
quantum state. How to interchange different resources has ever
been a question in quantum computation and quantum information. In
1993, Bennett {\it et al}[1] first presented a quantum
teleportation scheme. In the scheme, an arbitrary unknown quantum
state in a qubit can be teleported to a distant qubit with the aid
of Einstein-Podlsky-Rosen (EPR) pair. Their work showed in essence
the interchangeability of different resources in quantum
mechanics. Later, in 1998, Karlsson and Bourennane generalized
Bennett {\it et al}'s teleportation idea by using a 3-qubit
Greenberger-Horne-Zeilinger (GHZ) state $|000\rangle+|111\rangle$
instead of an EPR pair[2]. In their scheme, conditioned on one
receiver's measurement outcome, the other receiver can recover the
arbitrary quantum state initially in the sender's qubit. This
means different resources can be exchanged in a control manner. In
1999, Hillery, Buzek and Berthiaume [3] first and explicitly
presented the concept of quantum secret sharing, which is a
quantum version of classical sharing schemes [4]. In their work
not only classical messages but also quantum information can be
securely shared by two (three) parties via using three-particle
(four-particle) GHZ states. Since then, a number of works were
focused on quantum secret sharing of quantum information[5-10].
Very recently, Yang {\it et al}[11] have presented their extensive
study on teleporting multiqubit information with control of many
agents in a network. In their paper [11] they briefly reviewed how
to complete a controlled teleportation of multi-qubits by using
the methods in the Refs.[2-3] and then proposed their preponderant
method. Compared to the methods in the Refs.[2-3], as they claimed
in Ref.[11], their method is apparently simpler and economical,
because the required auxiliary qubit resources, the number of
local operations, and the quantity of classical communication are
greatly reduced. However, as the controlled teleportation is
concerned, we think, Yang {\it et al}'s method[11] is still very
complicated. In this paper, we will present a general idea to
construct various methods for multi-qubit teleportation between
two remote parties with control of many agents in a network. To
well demonstrate our idea, we will show three simple examples,
where the Ekert91 quantum key distribution (QKD) [12] and the
quantum secure direct communication (QSDC) protocol using single
photon [13] are employed to construct our methods. Our methods are
also able to successfully teleport multi-qubit quantum information
from a sender to a receiver with control of many agents, but in a
much simpler way. Moreover, the three simple examples will show
that, our methods are more feasible according to the present-day
technique, the required qubits resources and the preparation
difficulty of initial states are considerably reduced, and some
local operations are not needed at all. One will see these
advantages of our methods later.

Before demonstrating our idea, let us briefly review multi-qubit
teleportation. Suppose that Alice has a string of message qubits
labelled by $1,2,\dots,m$.  The initially state of these $m$
qubits is $\Pi_{i=1}^m (\alpha_i |0\rangle_i +\beta_i
|1\rangle_i)$, where $\alpha_i$ and $\beta_i$ are arbitrary
complex coefficients. Alice wants to send this $m$-qubit quantum
information to a remote receiver Bob via teleportation. To this
end, Alice first prepares $m$ photon pairs all in same Bell
states, say, $\phi^+=(|00\rangle+|11\rangle)/\sqrt{2}$. Then the
state of all the qubits in her lab is
\begin{eqnarray}
\Pi_{i=1}^m (\alpha_i |0\rangle_i +\beta_i
|1\rangle_i)\phi^+_{i'i''}=\frac{1}{\sqrt{2}}\Pi_{i=1}^m (\alpha_i
|0\rangle_i +\beta_i
|1\rangle_i)(|00\rangle_{i'i''}+|11\rangle_{i'i''}).
\end{eqnarray}
It can be rewritten as
\begin{eqnarray}
\frac{1}{2} \Pi_{i=1}^m [\phi^+_{ii'}(\alpha_i |0\rangle_{i''}
+\beta_i|1\rangle_{i''}) + \psi^+_{ii'} (\alpha_i
|1\rangle_{i''} +\beta_i|0\rangle_{i''})\nonumber \\
+ |\phi^-_{ii'} (\alpha_i |0\rangle_{i''} -\beta_i|1\rangle_{i''})
+ \psi^-\rangle_{ii'} (\alpha_i |1\rangle_{i''}
-\beta_i|0\rangle_{i''}],
\end{eqnarray}
where $\phi^-=(|00\rangle-|11\rangle)/\sqrt{2}$ and
$\psi^{\pm}=(|01\rangle\pm|10\rangle)/\sqrt{2}$ are Bell states.
Alice sends the string of all $i''$ qubits to Bob and performs
Bell-state measurement on each two-photon pair $\{i,i'\}$ in her
lab. Alice publishes her measurement outcome. Conditioned on
Alice's measuremtn outcome, say, $\phi^+_{ii'}$
($\psi^+_{ii'}$,$\psi^-_{ii'}$,$\phi^-_{ii'}$), Bob performs a
unitary operation $I=|0\rangle\langle 0|+|1\rangle\langle 1|$
($u_1=|0 \rangle\langle 1|+|1 \rangle\langle 0|$, $u_2=|0
\rangle\langle 0|-|1\rangle\langle 1|$, $u_3=|0\rangle\langle
1|-|1 \rangle\langle 0|$) to reconstruct the unknown state in
Bob's qubit $i''$. To fully reconstruct the unknown states in
Bob's qubits, it is necessary for Bob to correctly know Alice's
all Bell-state measurement outcomes, otherwise, he can not know
for each his qubit what unitary operation he should perform. In
this sense, the messages of Alice's Bell-state measurement
outcomes are in essence control parameters on Bob's correct
reconstructions. In Refs.[2,3,11], to transform Alice's control
parameters (i.e., the measurement outcomes) into each agent's
control parameter on Bob's reconstructions, the authors have
designed complicated methods, where Alice needs to prepare
complicated states (e.g., the GHZ states in Refs.[2,3] and the
states as the equations 2, 21 and 29 in Ref.[11]), to perform some
local operations (for an example, the Hadamard operation in
Ref.[11]) and to perform Bell state measurements.  The essential
purpose of their methods is first to let each agent own a control
parameter and then to let Bob be able to fully reconstruct the
unknown states if all the agents collaborate with him. In fact,
the task to let each agent own a control parameter can be achieved
in a much simpler way. Our general idea is that, both Alice and
each agent first securely share this agent's control parameter and
then Alice uses this control parameter to uniformly encrypt her
Bell-state measurement messages outcomes before her public
announcements. Hence, only all the agents collaborate with Bob can
the unknown states be fully reconstructed in Bob's qubits. Alice's
share of each agent's control parameter can be easily achieved by
using various quantum key distributions (QKD) [10,12,14-30] or
various quantum secure direct communications (QSDC) [31-34] or
their combinations. Obviously, during this sharing procedure the
security can be assured.

To well demonstrate our idea, we will show three simple examples.
The first example employs the Ekert91 QKD protocol to construct a
method. Incidentally, so far there are many QKD protocols
[12,14-30] and anyone of them can be employed to construct a
method. In 1991, Ekert showed that quantum entanglement can be
useful in sharing private key between two parties. Suppose that
Alice and Charlie share many maximally entangled pairs of qubits.
They then make measurements in jointly determined random bases.
After the measurements, Alice and Charlie publish which basis they
have used.  If they had used the same basis, then  the key would
be perfectly correlated. Instead of discarding the keys resulting
from the measurement in different bases, Alice and Charlie can use
them to check whether Bell's inequality is satisfied or not. If it
is, then the attacker Eve's presence is detected, otherwise, Eve
is absent and they can keep the perfect correlated keys. Hence one
can use the Ekert91 QKD protocol to securely generate a correlated
key between two parties. Suppose that Alice and Charlie share
securely a key '1'. Define that $\phi^+$,$\phi^-$,$\psi^+$ and
$\psi^-$ correspond to two classical messages '00','01','10' and
'11', respectively. After Alice's Bell state measurements, without
loss of generality, we suppose that she obtains $\{\psi^+_{11'},
\phi^-_{22'},\dots, \psi^-_{mm'}\}$ corresponding to classical
bits $\{(10)_{11'},(01)_{22'},\dots, (11)_{mm'}\}$. Alice first
uses the shared key '1' to encrypt her measurement outcomes
according to the definitions
$'00'+'0'='00','01'+'0'='01','10'+'0'='10','11'+'0'='11','00'+'1'='01','01'+'1'='10','10'+'1'='11','11'+'1'='00'$,
and then publicly announces the string
$\{(11)_{11'},(10)_{22'},\dots, (00)_{mm'}\}$. If Charlie
collaborates with Bob, i.e., he tells Bob the shared key is '1',
then Bob can extract the correct measurement outcomes from Alice's
public announcements and performs correct unitary operation on
each his qubit. This means that the unknown states can be fully
reconstructed in Bob's qubits.  Otherwise, Bob's reconstruction of
the unknown states can not be fully achieved. Above we have shown
a method of teleportation of unknown states with control of an
agent. If there is one more agent and it is Dick, then via the
Ekert91 QKD protocol Alice and Dick can also share securely a key,
say, '0', then Alice encrypt her measurement outcomes according to
the key '0' shared commonly by her and Dick after her first
encryption using the key '1' shared commonly by her and Charlie.
After her two uniform encryptions, Alice publicly announces the
classical bits '$\{(11)_{11'},(10)_{22'},\dots, (00)_{mm'}\}$'. If
Charlie and Dick collaborate with Bob, then Bob can correctly
extract Alice's original measurement outcomes and fully
reconstruct the unknown states. Otherwise, Bob's full
reconstructions fail. Similarly, one can find that a many agent
case can be easily achieved by using the Ekert91 QKD protocol many
times.

Our second example employs the quantum secure direct communication
(QSDC) protocol using single photons[13]. This protocol has been
proven to be unconditionally secure[13]. This means Charlie can
securely inform Alice to encrypt his control bit on Alice's
measurement outcomes before Alice's public announcements. Of
course, other agents can also take advantage of this QSDC protocol
to securely let Alice encrypt their control bits on Alice's
measurement outcomes. Only all the agents collaborate with Bob can
he first extract the messages of Alice's measurement outcomes and
then fully reconstruct the unknown states by performing correct
unitary operations. Otherwise, Bob's full reconstructions fail. By
the way, so far there are several QSDC protocols[13,31-34] and
anyone of them can replace the QSDC employed in the present paper
to establish a method.

Our third example employs both the Ekert91 QKD protocol and the
QSDC protocol using single photons. From our first two examples,
one knows that any agent can randomly choose one of the two
protocols to own a control parameter. Hence, it is obvious that
the combining use of the Ekert91 QKD protocol and the QSDC
protocol using single photons is also suitable.

By the way, another important method is that Alice herself
uniformly encrypts her measurement outcomes first and then lets
all the agents secretly share her encryption bit according to
quantum secret sharing, for so far many quantum secret sharing
schemes [36-43] have been proposed. Since we have mentioned such
important method in our previous paper [44], here we will not
introduce it anymore.

Let us show some comparisons. From our present methods and the
method in Refs.[2,3,11], one can see that, (1) The complicated
initial states needed to to be prepared, such as the
multi-particle GHZ states in Refs.[2,3] and the complicated
entangled states as the equations 2,21 and 29 in Ref.[11], are not
necessary in the present methods. Bell states and single-photon
states are sufficient for use in the present methods. Hence, the
present methods greatly reduces the required auxiliary qubit
resources and the preparation difficulty of initial states; (2) In
each of our methods only the Bell states are needed to be
identified by Alice whenever how many agents are. However,
according to the methods in Refs.[2,3,11], the identification of
multi-particle GHZ states should be completed by Alice when the
number of agents is not less than 2. It is obvious in the present
methods the difficulty of Alice's identification on her entangled
states is reduced. (3) To our knowledge, so far preparation of
five-photon entangled states has been achieved in experiment[45],
however, preparation more-photon entanglement is still desired.
Alternatively, when the number $n$ of the agents is large, it is
impossible to prepare $n$-photon GHZ states according to the
present-day technologies. Hence, in the case that $n$ is large,
the methods in Refs.[2,3,11] are all impossible in reality. In
contrast, according to the present methods, for any large number
of the agents, the efficient controls can be realized, for Bell
states and single-photon states are sufficient for use in the
present methods. Hence the present methods are more feasible
according to the present-day technique; (4) Local operations on
the agents' qubits such as the Hadamard operation in Ref.[11] are
not needed in the present methods at all.

To summarize, we have presented a general idea to construct
various methods for multi-qubit quantum information with control
of many agents in a network. Simple examples employing the Ekert91
QKD [12] and the QSDC protocol [13] to construct various methods
are given to demonstrate our idea. They show that, at least some
of our methods are more feasible according to the present-day
technique, the required auxiliary qubits resources and the
preparation difficulty of initial states are greatly reduced,
and some local operations are not needed at all. \\

\noindent {\bf Acknowledgements} \\
We thanks to Dr. X. B. Wang for his helps. This work is supported
by the National Natural Science Foundation of China under Grant No. 10304022. \\

\noindent {\bf References}

\noindent[1] C. H. Bennett, G. Brassard
C. Crepeau,  R. Jozsa, A. Peres and W. K. Wotters, Phys. Rev.
Lett. {\bf70}, 1895 (1993).

\noindent[2] A. Karlsson and M. Bourennane, Phys. Rev. A {\bf 58},
4394 (1998).

\noindent[3] M. Hillery, V. Buzk  and A. Berthiaume, Phys. Rev. A
{\bf 59}, 1829 (1999).

\noindent[4] A. Shamir, Commun. ACM {\bf 22}, 612 (1979).

\noindent[5] R. Cleve, D. Gottesman  and H. K. Lo, Phys. Rev.
Lett. {\bf 83}, 648 (1999).

\noindent[6] S. Bandyopadhyay, Phys. Rev. A {\bf 62}, 012308
(2000).

\noindent[7] L. Y. Hsu, Phys. Rev. A {\bf 68}, 022306 (2003).

\noindent[8] Y. M. Li, K. S. Zhang and K. C. Peng, Phys. Lett. A
{\bf 324}, 420  (2004).

\noindent[9] A. M. Lance, T. Symul, W. P. Bowen, B. C. Sanders and
P. K. Lam,  Phys. Rev. Lett. {\bf 92}, 177903  (2004).

\noindent[10] Z. J. Zhang, J. Yang, Z. X. Man and Y Li, Eur. Phys.
J. D {\bf 33} 133 (2005).

\noindent[11] Chui-Ping Yang, Shi-I Chu and Siyuan Han, Phys. Rev.
A {\bf 70}, 022329 (2004).

\noindent[12] A. K. Ekert, Phys. Rev. Lett. {\bf 67}, 661 (1991).

\noindent[13] F. G. Deng and G. L. Long,  Phys. Rev. A {\bf69},
 052319 (2004).

\noindent[14] C. H. Bennett, Phys. Rev. Lett. {\bf68}, 3121
 (1992).

\noindent[15] C. H. Bennett, G. Brassard, and N.D. Mermin, Phys.
Rev. Lett. {\bf68}, 557(1992).

\noindent[16] L. Goldenberg and L. Vaidman, Phys. Rev. Lett.
{\bf75}, 1239  (1995).

\noindent[17] B. Huttner, N. Imoto, N. Gisin, and T. Mor, Phys.
Rev. A {\bf51}, 1863 (1995).

\noindent[18] S. J. D. Phoenix, S. M. Barnett, P. D. Townsend, and
K. J. Blow, J. Mod. Opt. {\bf42}, 1155 (1995).

\noindent[19] M. Koashi and N. Imoto, Phys. Rev. Lett. {\bf79},
2383 (1997).

\noindent[20] W. Y. Hwang, I. G. Koh, and Y. D. Han, Phys. Lett. A
{\bf244}, 489 (1998).

\noindent[21] P. Xue, C. F. Li, and G. C. Guo,  Phys. Rev. A
{\bf65}, 022317 (2002).

\noindent[22] H. Bechmann-Pasquinucci and N. Gisin, Phys. Rev. A
{\bf59}, 4238 (1999).

\noindent[23] A. Cabello, Phys. Rev. A {\bf61},052312 (2000);
{\bf64}, 024301 (2001).

\noindent[23] A. Cabello, Phys. Rev. Lett. {\bf85}, 5635 (2000).

\noindent[25] G. P. Guo, C. F. Li, B. S. Shi, J. Li, and G. C.
Guo, Phys. Rev. A {\bf64}, 042301 (2001).

\noindent[26] G. L. Long and X. S. Liu, Phys. Rev. A {\bf65},
032302 (2002).

\noindent[27] F. G. Deng and G. L. Long, Phys. Rev. A {\bf68},
042315 (2003).

\noindent[28] J. W. Lee, E. K. Lee, Y. W. Chung, H. W. Lee, and J.
Kim, Phys. Rev. A {\bf 68}, 012324 (2003).

\noindent[29] Daegene Song, Phys. Rev. A {\bf69}, 034301 (2004).

\noindent[30] X. B. Wang, Phys. Rev. Lett. {\bf 92}, 077902
(2004).

\noindent[31] A. Beige, B. G. Englert, C. Kurtsiefer, and
H.Weinfurter, Acta Phys. Pol. A {\bf101}, 357 (2002).

\noindent[32] Kim Bostrom and Timo Felbinger, Phys. Rev. Lett.
{\bf89}, 187902 (2002).

\noindent[33] F. G. Deng, G. L. Long and  X. S. Liu, Phys. Rev. A
{\bf68},  042317 (2003).

\noindent[34] Z. J. Zhang, Z. X. Man and Yong Li, International
Journal of Quantum information {\bf 2}, 521 (2004).

\noindent[35] N. Gisin, G. Ribordy, W. Tittel, and H. Zbinden,
Rev. Mod. Phys. {\bf 74}, 145 (2002).

\noindent[36] A. Karlsson, M. Koashi and N. Imoto, Phys. Rev. A
{\bf 59}, 162  (1999).

\noindent[37] D. Gottesman,  Phys. Rev. A {\bf 61}, 042311
(2000).

\noindent[38] W. Tittel, H. Zbinden and N. Gisin, Phys. Rev. A
{\bf 63}, 042301  (2001).

\noindent[39] V. Karimipour and A. Bahraminasab, Phys. Rev. A {\bf
65}, 042320 (2002).

\noindent[40] H. F. Chau, Phys. Rev. A {\bf 66}, 060302  (2002).

\noindent[41] S. Bagherinezhad and V. Karimipour, Phys. Rev. A
{\bf 67}, 044302  (2003).

\noindent[42] G. P. Guo and G. C. Guo, Phys. Lett. A {\bf 310},
 247 (2003).

\noindent[43] L. Xiao, G. L. Long, F. G. Deng, and  J. W. Pan,
Phys. Rev. A {\bf 69}, 052307 (2004).

\noindent[44] Z. J. Zhang, Y. Li and Z. X. Man, Phys. Rev. A {\bf
71}, 044301 (2005).

\noindent[45] Zhi Zhao, Yu-Ao Chen, An-Ning Zhang, Tao Yang, Hans
J. Briegel, and Jian-Wei Pan, Nature (London), {\bf 430}, 54
(2004).

\enddocument